%% file: main.tex
\documentclass[conference]{IEEEtran}
\IEEEoverridecommandlockouts
\usepackage{amsmath,amssymb,amsfonts}
\usepackage{algorithmic}
\usepackage{graphicx}
\usepackage{textcomp}
\usepackage[table,xcdraw]{xcolor}
\usepackage{balance}
\usepackage{soul}
\usepackage{booktabs}
\usepackage{multirow}
\usepackage{float}
\usepackage{url}
\usepackage{tcolorbox}
\usepackage{tabularx} 
\usepackage[noadjust]{cite}
\usepackage[hidelinks,bookmarks=true,pagebackref=true]{hyperref}
\usepackage{subcaption}
\usepackage{xurl}
\usepackage{enumitem}
\setlist[itemize,enumerate]{noitemsep, topsep=0pt, leftmargin=1.0em}
\usepackage[utf8]{inputenc}
\usepackage{newunicodechar,graphicx}
\DeclareRobustCommand{\okina}{%
  \raisebox{\dimexpr\fontcharht\font`A-\height}{%
    \scalebox{0.8}{`}%
  }%
}
\newunicodechar{ʻ}{\okina}
\usepackage{background}
\backgroundsetup{
  position=current page.east,
  angle=-90,
  nodeanchor=east,
  vshift=-4mm,
  opacity=0.5,
  scale=3,
  contents=PPREPRINT
}
\newcommand{\rev}[1]{\textcolor{black}{#1}}

\newcommand{\RQA}{\textbf{RQ1}: What are the common practices and challenges faced by mobile app developers in securing their apps?}

\newcommand{\RQAa}{\textbf{RQ1.1}: What security features are frequently implemented by mobile app developers?}

\newcommand{\RQAb}{\textbf{RQ1.2}: What steps do mobile app developers take to ensure the security of their apps and third-party components?}

\newcommand{\RQAc}{\textbf{RQ1.3}: What are the technical and non-technical challenges faced by mobile app developers in securing their apps?}

\newcommand{\RQB}{\textbf{RQ2}: What resources do mobile app developers utilize for help with mobile app security topics?}

\newcommand{\RQBa}{\textbf{RQ2.1}: How do mobile app developers engage with Stack Overflow for help with mobile app security topics?}

\newcommand{\RQBb}{\textbf{RQ2.2}: What other resources do mobile app developers use for help with mobile app security topics?}

\newcommand{\RQC}{\textbf{RQ3}: To what extent do current learning materials prepare mobile app developers to build secure apps, and what guidance can experienced developers offer to improve mobile app security education and practices?}

\begin{document}

\title{A Developer-Centric Study Exploring Mobile Application Security Practices and Challenges}

\author{\IEEEauthorblockN{Anthony Peruma}
\IEEEauthorblockA{\textit{University of Hawaiʻi at Mānoa}\\
Hawaiʻi, USA \\
peruma@hawaii.edu}
\and
\IEEEauthorblockN{Timothy Huo}
\IEEEauthorblockA{\textit{University of Hawaiʻi at Mānoa}\\
Hawaiʻi, USA \\
thuo@hawaii.edu}
\and
\IEEEauthorblockN{Ana Catarina Araújo}
\IEEEauthorblockA{\textit{University of Hawaiʻi at Mānoa}\\
Hawaiʻi, USA \\
acoa@hawaii.edu}
\and
\IEEEauthorblockN{Jake Imanaka}
\IEEEauthorblockA{\textit{University of Hawaiʻi at Mānoa}\\
Hawaiʻi, USA \\
jimanaka@hawaii.edu}
\and
\IEEEauthorblockN{Rick Kazman}
\IEEEauthorblockA{\textit{University of Hawaiʻi at Mānoa}\\
Hawaiʻi, USA \\
kazman@hawaii.edu}
}

\maketitle

\begin{abstract}
Mobile applications (apps) have become an essential part of everyday life, offering convenient access to services such as banking, healthcare, and shopping. With these apps handling sensitive personal and financial data, ensuring their security is paramount. While previous research has explored mobile app developer practices, there is limited knowledge about the common practices and challenges that developers face in securing their apps. Our study addresses this need through a global survey of 137 experienced mobile app developers, providing a developer-centric view of mobile app security.

Our findings show that developers place high importance on security, frequently implementing features such as authentication and secure storage. They face challenges with managing vulnerabilities, permissions, and privacy concerns, and often rely on resources like Stack Overflow for help. Many developers find that existing learning materials do not adequately prepare them to build secure apps and provide recommendations, such as following best practices and integrating security at the beginning of the development process. We envision our findings leading to improved security practices, better-designed tools and resources, and more effective training programs.
\end{abstract}


\section{Introduction}

The mobile application (app) development landscape has evolved rapidly in recent years. Advancements in software tools and technologies, such as feature-rich IDEs like Android Studio, Xcode, and Visual Studio, the availability of various SDKs and libraries, and cross-platform frameworks such as React Native and Flutter have greatly enhanced the abilities of both experienced and novice developers to create mobile applications for diverse purposes across multiple platforms \cite{KOCH20141423,Puvvala2016,Koram2023}. This is clearly evident by the sheer number of apps available on app marketplace stores such as the Apple App Store and Google Play Store; as of Q3 2022, there were over 3 million and 1 million apps on the Google Play Store and Apple App Store, respectively \cite{AppStore_stats}.

While this abundance of apps has benefited consumers by providing access to a wide range of services and information, it also introduces potential risks to user privacy and security. As apps increasingly handle sensitive personal and financial data, ensuring robust protection of users' privacy and security becomes paramount for app developers \cite{Nema2022,Martínez2014}. 

In reality, this is not always the case. Studies show that apps, both free and paid, contain many vulnerabilities that make them susceptible to  security threats. These vulnerabilities can be attributed to reasons such as poor design and programming mistakes, lack of security and API knowledge, and the use of third-party libraries \cite{Watanabe2017,DiSorbo2021,Gao2021}. Moreover, the prevalence of these issues is not limited to obscure or less popular apps; even well-known and widely used apps have been found to contain significant security vulnerabilities. For example, the much-publicized 2014 vulnerability in the Starbucks iOS app saved user credentials in plain text, exposing users to potential data theft \cite{Online-Starbuck,CVE-2014-0647}. Similarly, in 2018, the dating app Tinder, with millions of users worldwide, was found to be transmitting user images in an unencrypted format, potentially allowing malicious actors to intercept and misuse personal photos \cite{Online-Tinder,CVE-2018-6017}. In recent news, Microsoft Threat Intelligence discovered a path traversal vulnerability in multiple popular Android applications hosted on Google Play. This vulnerability could enable a malicious application to overwrite files in the vulnerable application's home directory \cite{Online-Microsoft}.

Incidents like the above highlight the importance of supporting developers in securing their apps. As a result, the research community has investigated app security through multiple types of studies, such as mining software repositories to analyze artifacts like source code, reverse engineering distribution packages, and the implementation and running of security-specific tools \cite{Xu2017Survey,Ebrahimi2021Survey,Zhan2022,Li2017,Senanayake2023}. While empirical studies of source code and other related artifacts allow researchers to conduct extensive, reproducible studies with generalizable results, they also pose challenges. These approaches, while valuable, often fail to capture the full context of development decisions, the rationale behind certain security practices, and the real-world constraints faced by developers.

Although there are studies that survey app developers on their practices and challenges in securing their apps, research in this area remains limited. As we discuss in Section \ref{Section:related}, most existing studies either focus on a specific domain or technology, or security is not the primary topic of the survey. Further, the study by Balebako et al. \cite{Balebako2014}, which focuses on security challenges app developers face, was conducted almost a decade ago. Given the advancements in the field, there is a need to investigate the current landscape of app security.

\subsection*{Goal \& Research Questions}
\label{Section:Goal}
In this study we aim, to provide a more recent and \textit{developer-centric view of the security practices and challenges in app development}. To achieve this, we conducted a global online survey of 137 experienced mobile app developers. Through a set of quantitative and qualitative questions, our study identifies common practices and obstacles that developers face in securing their applications, investigates their use of resources for assistance with security topics, and assesses the adequacy of current learning materials. We envision our findings advancing the field of mobile app security by offering recent and real-world perspectives, leading to improvements in security practices, developer resources, tools, and security-focused education. Our study aims to address the following research questions (RQs):

\vspace{1mm}
\noindent\textbf{\RQA} The insights gained from this RQ lead us to better understand the decision-making processes of developers when it comes to implementing security features in their apps. This understanding will guide the development of more practical security guidelines, training programs, and tools.

\vspace{1mm}
\noindent\textbf{\RQB} This RQ investigates
the diversity of information sources influencing mobile app security practices and provides insight for effectively disseminating security information and best practices.

\vspace{1mm}
\noindent\textbf{\RQC} This RQ aids in understanding the effectiveness of learning materials in preparing app developers to address real-world security challenges. Further, insights from experienced developers offer practical guidance to enhance security education and practices.

\section{Related Work}
\label{Section:related}
As our study involves real-world developer challenges and practices on app security, here we only report on studies that survey/interview developers on mobile app security or examine discussions about app development on Stack Overflow.

\subsection{Surveys/Interviews on Mobile App Security}
Balebako et al. \cite{Balebako2014} conducted a study on the privacy and security behaviors of app developers. They interviewed 13 developers and surveyed 228 developers. The findings revealed that developers in small companies are more inclined to seek advice from their social network, while developers in larger companies are more likely to consult security specialists within their companies. Additionally, smaller companies are less likely to prioritize and implement robust privacy and security measures due to limited resources. The authors also reported that app developers tend to rely more on off-the-shelf or third-party tools for security rather than privacy. Furthermore, they have difficulties in reading and understanding the policies and terms of use of these libraries and are not always aware of the data collected by these libraries. \rev{While Balebako et al.'s study provided valuable insights, it was conducted almost a decade ago. Our study aims to provide a more current perspective on these issues and examine how these patterns may have changed with the evolution of the mobile app ecosystem.}

Acar et al. \cite{Acar2016} conducted a study with Android app developers and found that developers often use Stack Overflow to solve programming problems, including security issues. However, the study reported that Stack Overflow also contains insecure answers, so developers are cautious when using it. The study also observed that the official Android API documentation is more difficult to understand compared to Stack Overflow. \rev{Our work expands on this study by further examining the use of Stack Overflow and the other mobile app security resources developers utilize.}

In their survey with 97 developers of mHealth apps, Aljedaani et al. \cite{Aljedaani2020} report the following eight challenges in developing mHealth Apps: insufficient security knowledge, budget constraints, lack of security experts, poor security implementation decisions, time and cost constraints, lack of security testing, assumption of user disinterest in security, and legal and regulatory challenges. In their study, Ekambaranathan et al. \cite{Ekambaranathan2020} interviewed 20 Android developers working on family genre apps and found that developers often rely on third-party libraries from major companies, considering them as industry standards. However, developers expressed confusion regarding the functionality of these libraries and encountered challenges in understanding how data is managed by third-party libraries. \rev{Unlike the studies by Aljedaani et al. and Ekambaranathan et al., our study examines security challenges across a broader range of app types and is not limited to a specific domain.}

Francese et al. \cite{Francese2017} conducted a study involving software managers and professionals to explore crucial aspects of developing and managing mobile applications. They found that companies prioritize secure data transmission and storage, tailoring security strategies to customer requirements. Examples include using keychain in iOS, custom secured containers in Android, SSL with additional traffic encryption, HTTPS, local encryption, and complying with client IT policy for enterprise apps. Companies often have a dedicated security team and may partner for enterprise app security management, depending on the needed security level. \rev{Our study extends this work by identifying additional areas of concern and practices.} %They reported that companies prioritize secure data transmission and storage, often tailoring their security strategies to customer requirements. They use keychain in iOS and custom secured containers in Android, and adopt SSL with the option for additional traffic encryption. Social media apps use HTTPS and local encryption, while banking or payment apps enhance mobile security. Enterprise apps must comply with client IT policy, and companies typically have a dedicated security team with an assigned security champion developer. Depending on the needed security level, companies may partner for enterprise app security management. 

Although not security-focused, developer interviews and surveys by Joorabchi et al. \cite{Joorabchi2013} showed that most security testing is conducted manually. In a survey of app developers in Brazil, on mobile app testing practices, Santos et al. \cite{Santos2020} found that security testing is usually performed by QA analysts only. The authors also noted that challenges in testing are due to compatibility issues, lack of automation, and lower priority, with company culture being a primary reason for the lower priority for testing activities. \rev{Gardner et al. \cite{Gardner2022} survey iOS developers about privacy labels for their apps. The authors highlight the need for tools to assist developers in creating accurate privacy labels and complying with privacy regulations.}
While app security is not directly addressed, in a survey with 400 app developers, Tahaei et al. \cite{Tahaei2021} reported that some participants expressed concerns about the potential security risks associated with personalized ads.
\rev{Jorgensen et al. \cite{Jorgensen2015} investigated the risks associated with mobile applications through interviews with security experts and surveys of typical Android users. Their study identified common risks highlighted by both groups, such as concerns regarding the privacy of personal information, financial risks, and data integrity issues.}

\subsection{Stack Overflow Discussions on Mobile App Development}
In a study by Tahaei et al. \cite{Tahaei2022EuroUSEC}, 269 Stack Overflow posts about privacy and permissions in health apps for Android and iOS were analyzed. The study identified themes including developer confusion about privacy requirements, concerns about third-party tools, and the importance of developing tools for detecting health data usage and privacy compliance checks.
Fischer et al. \cite{Fischer2017SP} studied the effect of copying code from Stack Overflow on the security of Android apps. They found that many apps on the Google Play Store include insecure code snippets from Stack Overflow. Beyer and Pinzger \cite{beyer2014manual} examined 450 Android-related posts to understand the main issues and topics of Android app developers. They found that security-related posts were the least common, but the authors did not analyze these posts. In their general study of mobile app discussions, Rosen and Shihab \cite{Rosen2016MobileDevelopers} found that AIP-related questions are a common challenge for mobile app developers, but surprisingly, security is not among the discussed topics. Similarly, Linares-V\'{a}squez et al. \cite{Linares2014MSR} do not list security as a primary topic concerning mobile app development.\rev{Our study, in contrast to the above-mentioned studies, explores the effectiveness of using Stack Overflow for app security assistance and is not limited to a specific application domain.}

\rev{
\subsection{Summary} 
While prior research shows that security is an essential area in mobile app development, there is scope to better understand the real-world challenges and practices of app developers in securing their apps. Our study differentiates itself by providing a more comprehensive, global, and up-to-date view of mobile app security practices. By surveying 137 experienced mobile app developers from 23 countries, it offers a broader, developer-centric approach that explores general security practices and gathers practical recommendations to improve security practices and education.
}

\section{Method}
\label{Section:experiment_design}
This section outlines our survey design, participant recruitment, and the approach to analyzing the survey responses. Since our research involves human subjects, we took the necessary steps to ensure ethical compliance. Before publishing our survey and commencing data collection, we sought and received approval from the Institutional Review Board (IRB) of the Office of Research Compliance at our institution.

\subsection{Survey Design}
We utilized Qualtrics \cite{Qualtrics} to construct and host the survey and configured it to allow only one response per participant. Our survey comprises 24 questions designed to examine the participant demographics, experience in programming and mobile app development, their approach towards securing mobile apps, the challenges they face, resources they rely on for addressing app security issues, and any general feedback they wish to provide. As per best practices \cite{Kitchenham2002,kasunic2005designing,linaaker2015guidelines}, we formulated these questions based on the objectives of our study, as discussed in Section \ref{Section:Goal}, and our review of relevant literature, as discussed in Section \ref{Section:related}. Table \ref{Table:SurveyQuestions} displays the questions in the survey, the question type, whether the question requires a response, and logic/notes, if any. The complete questionnaire, as shown to participants, is available at: \cite{ProjectWebsite}.

\input{table_questionnaire}

\subsection{Survey Participants}
We used LinkedIn, a large professional social network, to recruit participants for our study, resulting in a diverse and skilled pool \cite{LinkedInESEM}. To find potential participants, we searched for ``mobile developer,'' ``mobile software engineer,'' ``android developer,'' ``android software engineer,'' ``iOS developer,'' and ``iOS software engineer'' on LinkedIn, which resulted in around 450,000 results\footnote{At the time of conducting the study, LinkedIn search only provides an estimation/approximation of the number of search results.}\footnote{Android and iOS are the leading mobile operating systems \cite{MobileOS_market}.}. This method of purposive sampling allowed us to carefully identify our target population of mobile app developers with experience in Android or iOS development, ensuring that their responses would best address our research questions \cite{Torchiano2017,Baltes2022}. We contacted 100 developers associated with each of the six search terms, totaling 600 potential participants\footnote{Due to budget constraints, we could not use LinkedIn's premium version. As a result, we had limitations on the number of searches and connections we could make within a given period.}. 

We conducted a manual review of each potential participant's profile to ensure they had mobile app development experience. This involved \rev{the authors} examining the individual's employment history by reviewing their job titles and descriptions for instances indicating mobile app development experience. \rev{Profiles that indicated that the subjects had only  taken courses on security were excluded, as we aimed to recruit participants with practical experience.}  Each potential participant received an invitation (via LinkedIn Messaging) to participate in our study and a link to our online Qualtrics survey. Before taking the survey, participants were asked to review an informed consent document outlining the study's purpose, procedures, risks, and benefits and to agree to it before proceeding to the survey questions. Participants did not receive compensation for taking part in the survey.

\subsection{Pilot Study}
We followed best practices by conducting a pilot study before officially launching the survey \cite{linaaker2015guidelines}. \rev{The pilot study is an essential step in the survey process that helps to assess the validity of the survey instrument \cite{MOLLERI2020}.}
The pilot study involved working with ten participants with experience in mobile app development \rev{who were recruited from the authors' professional network.} We worked with them in multiple iterations to identify areas for refinement in our questionnaire. \rev{They were instructed to review the survey questionnaire carefully and provide unbiased and constructive feedback.} Insights gathered from the pilot study helped us identify questions that lacked clarity, required reordering, or necessitated a shift in answer type (for example, from single-choice to multi-choice). After the pilot study, the questionnaire was finalized and submitted to the IRB for approval. Once approval was obtained, the questionnaire was made publicly available.

\subsection{Data Analysis}
We used both quantitative and qualitative methods to analyze the survey data \cite{Wagner2020}. \rev{In our quantitative analysis, we utilized standard statistical techniques to summarize and present data from closed-ended questions. For the qualitative analysis of open-ended responses, we utilized a systematic thematic analysis approach. This involved two authors independently reviewing all responses, generating initial codes, and identifying key concepts and patterns. Next, to ensure reliability and reduce bias, the authors compared, discussed, and refined the codes to arrive at a finalized set of themes. When reporting our results in Section \ref{Section:experiment_results}, we specify the data analysis method employed to answer each RQ.}

% The quantitative analysis involved standard statistical techniques, and for the qualitative analysis, we reviewed participants' open-ended responses to identify common themes. Two of the authors independently reviewed and categorized the responses to ensure reliability, discussing and reaching an agreement on the final categories. %More details on specific techniques will be provided when answering our research questions.

\section{Results}
\label{Section:experiment_results}
This section presents our RQ results\footnote{Due to space constraints, in some parts of the writeup, we only report on the frequent observations. The complete breakdown is in our dataset at \cite{ProjectWebsite}.}. We should note that the responses from pilot participants were only used to identify issues with the questionnaire and are not included in our RQ results.
Before answering our RQs, we first report on the number of responses received and participant demographics.

\subsection*{\textbf{Survey Responses}}
The survey was open to the public from early May 2023 until early September 2023 and received a total of 166 responses during this period. However, not all participants answered all the questions. To maintain consistency in our results analysis, we only considered participants who consented to participate in the study (survey question \#1) and answered all the required questions, resulting in 137 valid responses. Therefore, our results are based solely on these 137 responses.

\subsection*{\textbf{Participant Demographics}}
To gain a better understanding of our survey participants, we collected demographic information through survey questions \#2 to \#9. This information included their current employment status, years of experience in general programming and mobile app development, the extent to which building mobile apps is part of their job, the number of mobile apps they have developed, and the mobile app distribution platforms they use. We placed these demographic questions at the beginning of the survey following recommended practices \cite{kasunic2005designing}. These fact-based questions are typically easier for participants to answer compared to more in-depth subsequent questions.

Starting with survey question \#3, most participants (127 or 92.70\%) described themselves as employed, with 116 having full-time employment. Notably, none of the participants were full-time students. Moving on, in terms of general programming experience (survey question \#4), 45 participants (or 32.85\%) have 3 to 5 years of experience, while 44 participants (32.12\%) and 34 participants (24.82\%) have between 6 to 10 and more than 10 years of experience, respectively. Moreover, 63 participants (or 46.0\%) have 3-5 years of mobile app development experience (survey question \#5), while 32 (23.40\%) have 6–10 years of experience. Only 6 participants have less than one year of mobile app development experience. 

\begin{figure}
    \centering
    \includegraphics[width=1\linewidth]{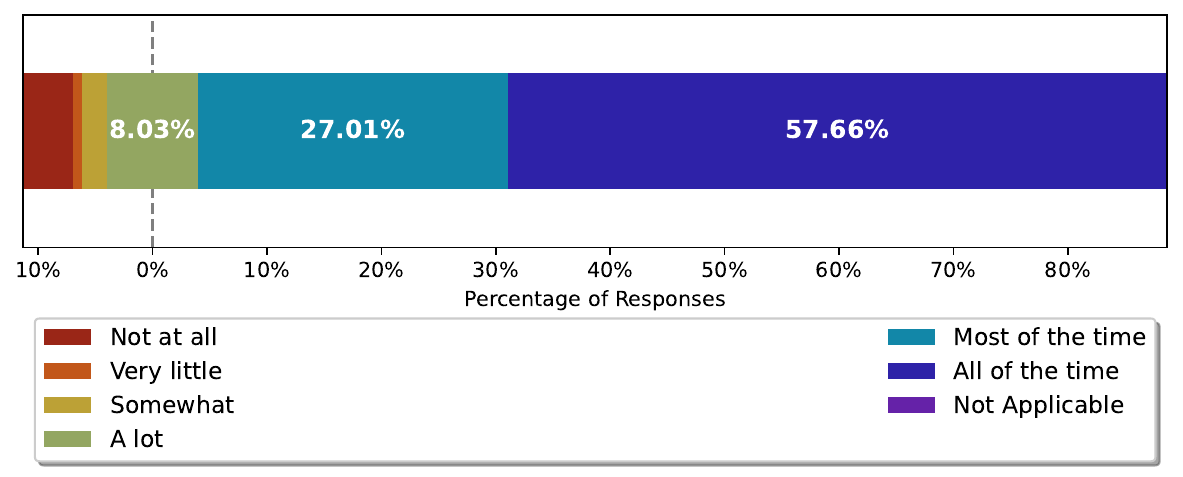}
    \caption{Extent of involvement in mobile app development.}
    \label{Figure:demo_6_likert}
\end{figure}

Further, as shown in Figure \ref{Figure:demo_6_likert}, 79 participants (or 57.66\%) answered ``All of the time'' when asked about the extent to which building mobile apps is part of their job/employment duties (survey question \#6), while 37 (27.01\%) answered ``Most of the time'' to the question. Filtering on these results, 107 full-time employed participants answered ``A lot,'' ``All of the time,'' or ``Most of the time'' to this question. 

In response to survey question \#7 asking about their involvement in building apps either as part of their job or as a hobby, 42 participants (30.66\%) stated they had been involved in building 3 to 6 apps. Additionally, 33 participants (24.09\%) and 31 participants (22.63\%) stated that they had been involved in building 6 to 10 apps and more than 10 apps, respectively. Next, to understand how the participants distribute their apps, we asked which app distribution platforms they utilize (survey question \#8). This question allowed for multiple choices, and 62 participants selected more than one platform. On average, respondents selected 1.49 platforms for this question. The most common pairing of app distribution platforms was the Apple App Store and Google Play, selected by 54 participants. Looking at each platform, we observe that 101 participants selected the Apple App Store, followed by 87 for Google Play. Other answers included Amazon Appstore, Firebase, Huawei App Gallery, and Microsoft Store.

The final question in our demographic section asks participants to identify their \textit{primary source} of mobile app development education (survey question \#9). Of the 137 participants, 77 (56.20\%) were self-taught (i.e., books/podcasts/videos/blogs), 18 (13.14\%) took an online course (e.g., edX, Udemy, Pluralsight, etc.), 17 (12.41\%) received formal education (i.e., college/university), 14 from a coding bootcamp/workshop (10.22\%). The ``Other'' was selected by 11 (8./03\%), out of which 5 mentioned that they learned about mobile app development through on-the-job training at their place of employment.

Finally, based on Qualtrics metadata, our results include participants from 22 countries. The United States had the highest number of participants at 37.23\%, followed by Brazil at 21.17\% and Argentina at 6.57\%.

Based on the demographic results, it is evident that our survey participants are experienced mobile app developers, increasing the likelihood of capturing real-world experiences, practices, and challenges of securing mobile apps.

%(Q10, Q11, Q12, Q13, Q14, Q15)
\subsection*{\RQA}
This RQ aims to provide an overview of how developers approach security in their mobile apps, the security features they commonly implement, and the obstacles they face. We answer this RQ through three specialized sub-RQs.

% Q10, Q11
\subsubsection*{\RQAa}
We answer this sub-RQ by analyzing the responses to survey questions \#10 and \#11. First, we examine the importance that app developers place on securing their apps, and then we look at the security features they frequently incorporate in their apps.

\begin{figure}
    \centering
    \includegraphics[width=1\linewidth]{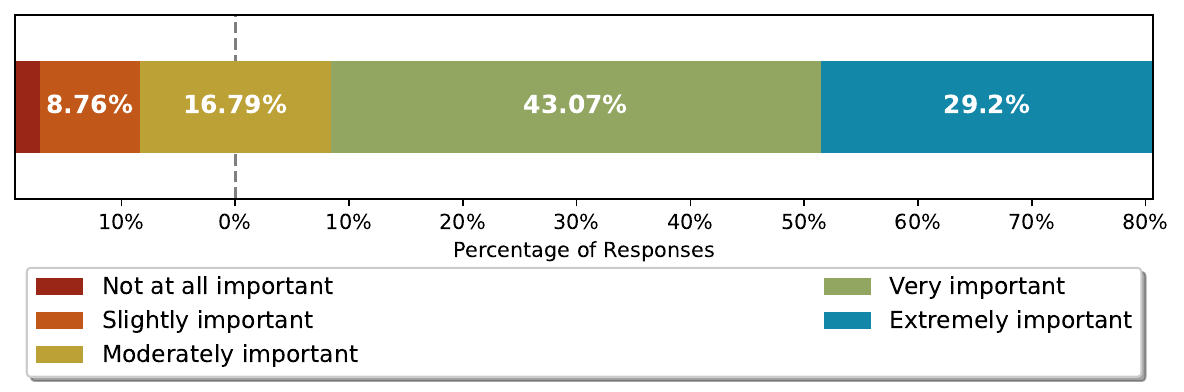}
    \caption{Importance of security during app development.}
    \label{Figure:demo_10_likert}
\end{figure}

To assess the importance of app security, participants used a Likert scale to rate its importance (survey question \#10). As shown in Figure \ref{Figure:demo_10_likert}, from the 137 participants, 59 participants (43.07\%)  answered ``Very important'', while 40 (29.20\%) answered ``Extremely important'', 23 (16.79\%) answered ``Moderately important'', and 12 (8.76\%) and 3 (2.19\%) answered ``Slightly important'' and ``Not at all important'', respectively. 

Next, participants were presented with a multi-choice question (survey question \#11) asking what security features they frequently implement in their mobile apps. As shown in Table \ref{Table:rq1-q11}, a total of 115 participants reported that authentication is a security feature they frequently implement, followed by permissions (105 participants) and secure storage of information (93 participants). Furthermore, the ``Other'' option was selected by 8 participants and included features such as ``debug detection'', ``simulator detection'', and ``Not sending private data when possible.''

As this is a multiple-choice question, 128 participants selected two or more answer options. On average, participants selected 4.68 answer options.
The most common two-answer combinations participants select are ``Authentication'' and ``Permissions'', occurring 96 times, followed by ``Authentication'' and ``Secure storage of information'' occurring 83 times. Looking at the most common three-answer combination, we observe ``Authentication'', ``Permissions'' and ``Secure storage of information'' occurring 69 times.

\begin{table}
\centering
\caption{Commonly implemented security features in apps.}
\vspace{-2mm}
\label{Table:rq1-q11}
\resizebox{\columnwidth}{!}{%
\begin{tabular}{@{}lrr@{}}
\toprule
\multicolumn{1}{c}{\textbf{Answer Option}}  & \multicolumn{1}{c}{\textbf{Count}} & \multicolumn{1}{c}{\textbf{Percentage}} \\ \midrule
Authentication                & 115 & 17.94\% \\ \midrule
Permissions                   & 105 & 16.38\% \\ \midrule
Secure storage of information & 93  & 14.51\% \\ \midrule
Data encryption               & 86  & 13.42\% \\ \midrule
Secure communications         & 71  & 11.08\% \\ \midrule
Biometric authentication      & 64  & 9.98\%  \\ \midrule
App hardening/obfuscation     & 54  & 8.42\%  \\ \midrule
Two-factor authentication     & 45  & 7.02\%  \\ \midrule
Other                         & 8   & 1.25\%  \\ \midrule
\multicolumn{1}{r}{\textit{\textbf{Total}}} & \textit{\textbf{641}}              & \textit{\textbf{100.00\%}}              \\ \bottomrule
\end{tabular}%
}
\end{table}

% Q12, Q13
\subsubsection*{\RQAb}
In this sub-RQ, we examine the practices app developers employ to secure their apps. Survey question \#12 examines general practices, while question \#13 centers on vulnerabilities resulting from the use of external dependencies. We include question \#13 because research shows that most app vulnerabilities stem from the use of third-party libraries \cite{Watanabe2017,Vargas2020}.

As shown in Table \ref{Table:rq1-q12}, when asked about the steps they take to ensure their apps are not vulnerable to common mobile app security risks (survey question \#12), there were 115 instances of participants adhering to secure coding practices, 57 instances of testing apps using penetration and other security-related testing tools, and 48 instances of conducting regular security audits or reviews. Additionally, 62 participants reported using two or more of these techniques, with an average of 1.66 answer options selected per participant. The most common two-answer combinations are ``Adherence to secure coding practices'' and ``Test the app using penetration and other security-related testing tools'', occurring 47 times, followed by ``Adherence to secure coding practices'' and ``Conduct regular security audits or reviews'', occurring 39 times. Examining the ``Other'' free-text responses, we observe participants mentioning using their party audits, static application security testing, and runtime application self-protection.

\begin{table}
\centering
\caption{Common practices for secure mobile apps.}
\vspace{-2mm}
\label{Table:rq1-q12}
\resizebox{\columnwidth}{!}{%
\begin{tabular}{@{}p{0.6\linewidth}rr@{}} 
\toprule
\multicolumn{1}{c}{\textbf{Answer Option}}                              & \multicolumn{1}{c}{\textbf{Count}} & \multicolumn{1}{c}{\textbf{Percentage}} \\ \midrule
Adherence to secure coding practices       & 115 & 50.66\% \\ \midrule
Test the app using penetration and other security-related testing tools & 57                                 & 25.11\%                                 \\ \midrule
Conduct regular security audits or reviews & 48  & 21.15\% \\ \midrule
Other                                      & 7   & 3.08\%  \\ \midrule
\multicolumn{1}{r}{\textit{\textbf{Total}}}                             & \textit{\textbf{227}}              & \textit{\textbf{100.00\%}}              \\
\bottomrule
\end{tabular}%
}
\end{table}

\begin{table}
\centering
\caption{Common practices for safeguarding against vulnerabilities in third-party dependencies.}
\vspace{-2mm}
\label{Table:rq1-q13}
\resizebox{\columnwidth}{!}{%
\begin{tabular}{@{}p{0.6\linewidth}rr@{}} 
\toprule
\multicolumn{1}{c}{\textbf{Answer Option}}                              & \multicolumn{1}{c}{\textbf{Count}}     & \multicolumn{1}{c}{\textbf{Percentage}} \\ \midrule
Using only well-established and trusted libraries/frameworks     & 106                  & 39.26\%              \\ \midrule
Regularly updating libraries/frameworks to their latest versions & 92                   & 34.07\%              \\ \midrule
Using vulnerability/security scanners                            & 33                   & 12.22\%              \\ \midrule
Rely on security reviews conducted by others on the libraries/framework & 30                                     & 11.11\%                                 \\ \midrule
Other                                                            & 9                    & 3.33\%               \\ \midrule
\multicolumn{1}{r}{\textit{\textbf{Total}}}                             & \textit{\textbf{270}}                  & \textit{\textbf{100.00\%}}              \\
\bottomrule
\end{tabular}%
}
\end{table}

\begin{table}
\centering
\caption{Frequent non-technical and technical challenges developers face in securing apps.}
\vspace{-2mm}
\label{Table:rq1-q14q15}
\resizebox{\columnwidth}{!}{%
\begin{tabular}{@{}p{0.6\linewidth}rr@{}} 
\toprule
\multicolumn{1}{c}{\textbf{Answer Option}} & \multicolumn{1}{c}{\textbf{Count}} & \multicolumn{1}{c}{\textbf{Percentage}} \\ \midrule
\multicolumn{3}{c}{\textit{Non-Technical Challenges}}                                                                         \\ \midrule
Limited access to security resources (e.g., personnel)     & 79                    & 24.16\%                    \\ \midrule
Balancing security with user experience, performance, and functionality  & 66                    & 20.18\%                    \\ \midrule
Keeping up with the latest security threats and vulnerabilities          & 65                    & 19.88\%                    \\ \midrule
Lack of awareness or understanding of mobile app security best practices & 58                    & 17.74\%                    \\ \midrule
Meeting compliance and regulatory requirements                           & 53                    & 16.21\%                    \\ \midrule
Other                                                                    & 6                     & 1.83\%                     \\ \midrule
\multicolumn{1}{r}{\textit{\textbf{Total}}}                              & \textit{\textbf{327}} & \textit{\textbf{100.00\%}} \\ \bottomrule
\multicolumn{3}{c}{\textit{Technical Challenges}}                                                                             \\ \midrule
Addressing vulnerabilities in third-party libraries and frameworks       & 67                    & 17.72\%                    \\ \midrule
Managing app permissions and user privacy concerns                       & 66                    & 17.46\%                    \\ \midrule
Protecting against reverse engineering                                   & 57                    & 15.08\%                    \\ \midrule
Preventing unauthorized access                                           & 53                    & 14.02\%                    \\ \midrule
Implementing secure data storage                                         & 52                    & 13.76\%                    \\ \midrule
Implementing secure communication protocols                              & 49                    & 12.96\%                    \\ \midrule
Implementing biometric authentication                                    & 25                    & 6.61\%                     \\ \midrule
Other                                                                    & 9                     & 2.38\%                     \\ \midrule
\multicolumn{1}{r}{\textit{\textbf{Total}}}                              & \textit{\textbf{378}} & \textit{\textbf{100.00\%}} \\ \bottomrule
\end{tabular}%
}
\end{table}

Moving on, as shown in Table \ref{Table:rq1-q13}, in terms of safeguarding against vulnerabilities of third-party libraries and frameworks (survey question \#13), there are 106 instances of participants indicating they use only well-established and trusted libraries/frameworks. Additionally, 92 participants regularly update libraries/frameworks to their latest versions, 33 use vulnerability/security scanners, and 30 rely on security reviews conducted by others on the libraries/frameworks. Furthermore, 88 participants reported using two or more techniques to protect against vulnerabilities, with an average of 1.97 answer choices. The most common two-answer combinations are ``Regularly updating libraries/frameworks to their latest versions'' and ``Using only well-established and trusted libraries/frameworks'', occurring 74 times, followed by ``Using only well-established and trusted libraries/frameworks'' and ``Using vulnerability/security scanners'', occurring 27 times. Finally, some participants used the ``Other'' option to indicate that they do not use third-party libraries.

% Q14, Q15
\subsubsection*{\RQAc}
This sub-RQ focuses on the typical technical and non-technical obstacles developers encounter in securing their apps, as addressed in survey questions \#14 and \#15. Table \ref{Table:rq1-q14q15} shows the frequent non-technical and technical challenges the participants encounter when securing their apps.

First, focusing on the non-technical challenges, the results reveal that resource constraints, such as personnel emerged as the most pressing issue to app security efforts for 79 participants. Following this, 66 participants had challenges incorporating security without compromising app performance or functionality, followed closely by challenges in staying abreast of rapidly evolving security threats and vulnerabilities. As this is a multi-choice question, 100 participants selected two or more challenges, with an average of 2.39 answer choices. The most common two-answer combinations are ``Balancing security with user experience, performance, and functionality'' and ``Limited access to security resources (e.g., personnel)'', occurring 38 times.

The main technical concern among participants was addressing vulnerabilities in third-party libraries and frameworks, with 67 participants citing it as a top issue. Managing app permissions and user privacy concerns was closely followed, with 66 participants expressing concerns about this. These two issues are the most prevalent challenges and are of nearly equal importance. Protecting against reverse engineering emerged as the third most significant challenge, with 57 participants indicating developers' struggles with safeguarding their app's intellectual property and sensitive algorithms. Similar to the non-technical challenges, 103 participants selected two or more technical challenges, with an average of 2.76 answer choices. The most common two-answer combinations are ``Implementing secure data storage'' and ``Managing app permissions and user privacy concerns'', occurring 37 times. This was followed by ``Preventing unauthorized access'' and ``Protecting against reverse engineering'', occurring 32 times. 

Most of the ``Other'' instances for both types of challenges are respondents stating they do not face challenges.

\begin{tcolorbox}[top=0.5pt,bottom=0.5pt,left=1pt,right=1pt]
\textbf{RQ1 Summary.}
Mobile app developers generally consider security very important, commonly implementing features like authentication, permissions, and secure storage. They typically adhere to secure coding practices, use trusted security tools and libraries, and regularly update their libraries. However, they face technical challenges such as managing vulnerabilities in third-party components, handling permissions and privacy concerns, and protecting against reverse engineering. Non-technical challenges include security resource constraints, balancing security with functionality and performance, and keeping up with evolving threats.
\end{tcolorbox}

%(Q16, Q17, Q18, Q19, Q20, Q21, Q22)
\subsection*{\RQB}
Building upon the insights gained from the previous RQ regarding developer practices in securing their apps and the challenges they encounter, this RQ seeks to understand the specific methods and resources developers rely on for seeking help and guidance on mobile app security topics.

%(Q16, Q17, Q18, Q19, Q20)
\subsubsection*{\RQBa}
In the first sub-RQ, we focus on the extent to which developers rely on Stack Overflow for assistance with mobile app security topics. We focus on Stack Overflow since it is the largest online programming question-and-answer platform with millions of questions, answers, and users \cite{SO_QuestionsUsers}. Additionally, past research has examined developer discussions on a variety of software engineering topics, including general mobile app development \cite{Rosen2016MobileDevelopers,Linares2014MSR,beyer2014manual,Fontao2017Emotions} and app security \cite{Tahaei2022EuroUSEC,Fischer2017SP}.   

\begin{figure}
    \centering
    \includegraphics[width=1\linewidth]{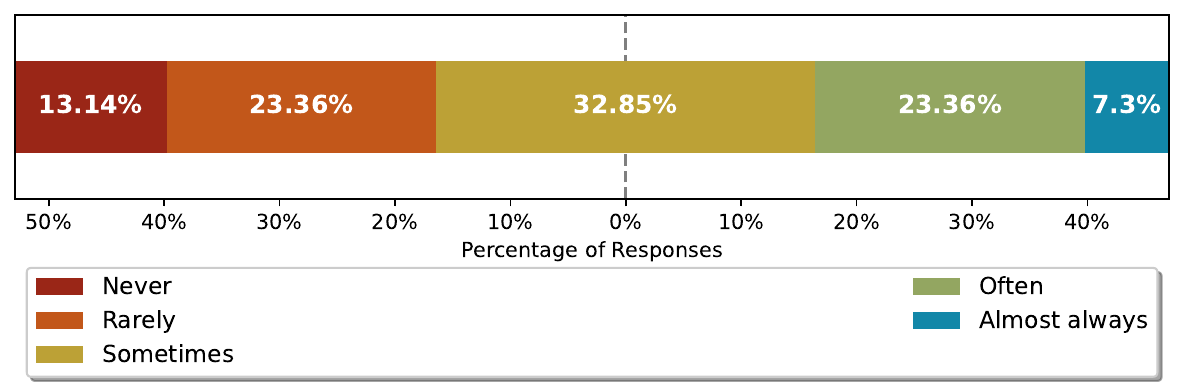}
    \caption{Stack Overflow usage for app security help.}
    \label{Figure:demo_16(17)_likert}
\end{figure}

Starting off with survey question \#16, we ask participants how frequently they utilize Stack Overflow for help with securing their apps. As shown in Figure \ref{Figure:demo_16(17)_likert}, the most common response was ``Sometimes'', selected by 45 participants which represent 32.85\% of the total. The next two most frequent answers were ``Often'' and ``Rarely'', each chosen by 32 participants or 23.36\% of respondents. Additionally, 18 participants (13.14\%), indicated that they ``Never'' use Stack Overflow for mobile app security help. These results indicate that while Stack Overflow is an important resource for many, its usage varies among developers, with a small minority not finding it helpful for their mobile app security needs.

\begin{figure}
    \centering
    \includegraphics[width=1\linewidth]{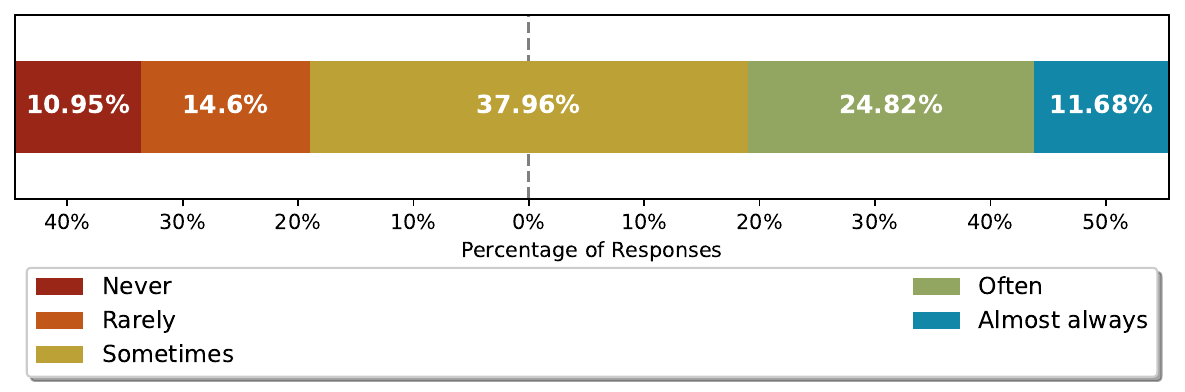}
    \caption{Finding helpful security information on Stack Overflow.}
    \label{Figure:demo_17(18)_likert}
\end{figure}

Moving on, in survey question \#17, we ask the participants the extent to which they find helpful security information on Stack Overflow. From Figure \ref{Figure:demo_17(18)_likert}, we observe 52 participants (37.96\%) reporting ``Sometimes'', 34 (24.82\%) reporting ``Often'', and 16 (11.68\%) reporting ``Almost always''. However, 20 participants (14.60\%) said they "Rarely" find helpful information, and 15 (10.95\%) said they "Never" do. These findings indicate that a majority of participants find Stack Overflow useful for obtaining helpful mobile app security content.

\begin{figure}
    \centering
    \includegraphics[width=1\linewidth]{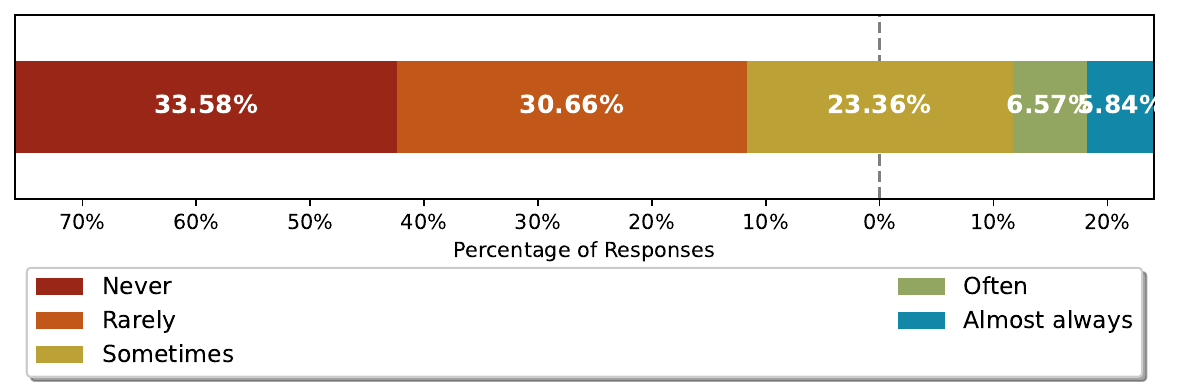}
    \caption{Frequency of asking questions on Stack Overflow.}
    \label{Figure:demo_18(19)_likert}
\end{figure}

\begin{figure}
    \centering
    \includegraphics[width=1\linewidth]{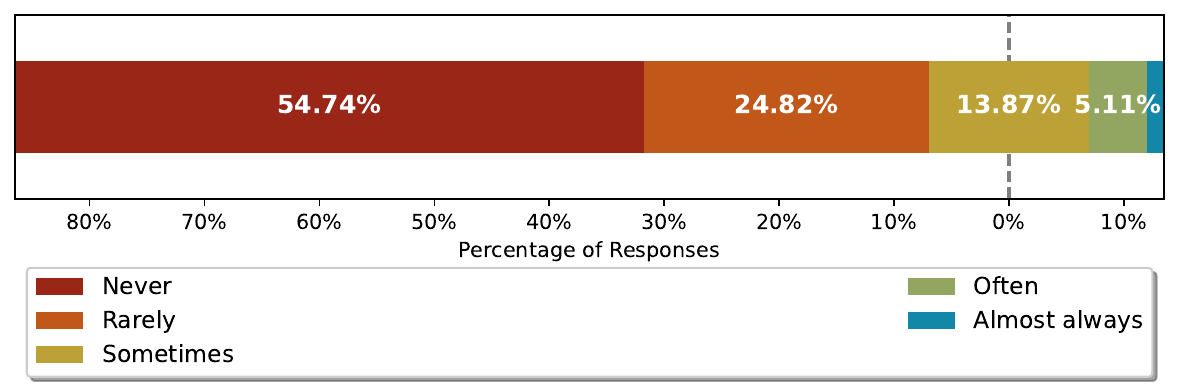}
    \caption{Frequency of answering questions on Stack Overflow.}
    \label{Figure:demo_19(20)_likert}
\end{figure}

Finally, survey questions \#18 and \#19, examine the extent to which developers ask and answer questions on Stack Overflow, respectively.
From Figure \ref{Figure:demo_18(19)_likert} and \ref{Figure:demo_19(20)_likert}, we observe that app developers are more likely to use Stack Overflow to find answers than to ask or provide them. Our findings show that 46 participants (33.58\% ) ``Never'' ask questions, and 42 (30.66\%) ``Rarely'' do so. Similarly, 75 participants (54.74\%) ``Never'' answer questions, and 34 (24.82\%) ``Rarely'' do. In contrast, only a small minority frequently engage in asking or answering questions. Specifically, 9 participants (6.57\%) ``Often'' ask questions, while  7 (5.11\%) ``Often'' answer them. Even fewer, 9 participants (5.84\%) ``Almost always''' ask questions, and just 2 (1.46\%) ``Almost always'' answer them.

%(Q21, Q22)
\subsubsection*{\RQBb}
We answer this sub-RQ using survey question \#20, an optional question that accepts free-text responses.

This question received 64 responses. Two authors manually reviewed the responses and determined the following categories through review and discussion:
\begin{itemize}
    \item \textbf{Official Documentation.} Developers consult the official documentation for mobile operating systems like Google Android and Apple iOS, as well as from security vendors like DexGuard. They also gain knowledge from events organized by product vendors, such as Apple's WWDC.
    \item \textbf{Online Forums.} Developers also access online forums for security help and discussions. The participants do not mention the names of the forums they visit; they use generic terms such as ``\textit{forum}'' or ``\textit{Apple Dev forums}''. 
    \item \textbf{Online Articles, Videos, and Blogs.} Developers also get security help from articles (such as those on ``Medium'') and blog posts written by industry experts and experienced developers to stay informed about mobile security topics and best practices. Some specific examples participants mention include ``\textit{YouTube}'', ``\textit{GeeksforGeeks}'', ``\textit{Computerworld}'', and blogs by ``\textit{NowSecure}'' and ``\textit{SANS Institute}''.  
    \item \textbf{Security-Specific Documentation.} Participants also mentioned referring to specific security standards documentation, such as the Open Web Application Security Project, PCI, and penetration testing reports.
    \item \textbf{Generative AI Chatbot.} Developers are leveraging AI-powered tools like ``\textit{ChatGPT}'' to get help with mobile security questions and issues.
    \item  \textbf{Books \& Research Publications.} Some developers find that traditional resources such as books and research papers (e.g., ``\textit{IEEE documents}'') are valuable tools for learning about mobile security concepts and best practices.
    \item  \textbf{Internal Resources.} Developers working in organizations often have access to internal resources that can assist their work, such as documentation and expertise from security teams or experienced colleagues.   
\end{itemize}

\noindent Though not an actual resource, participants also mention that they use web search engines (e.g., ``\textit{Google searches}'') to find information and solutions related to mobile security topics. Additionally, as these are free-text responses, some participants provided multiple resources in their responses, like: ``\textit{PCI documentation, OWASP documentation, ChatGPT}''.

\begin{tcolorbox}[top=0.5pt,bottom=0.5pt,left=1pt,right=1pt]
\textbf{RQ2 Summary.}
Most developers use Stack Overflow for mobile app security help and generally find the information helpful. They are more likely to seek answers than ask or answer questions on the platform. Beyond Stack Overflow, developers rely on a range of resources including official vendor documentation, online forums, articles and blogs, security-specific documentation, AI chatbots, books, research papers, and internal company resources.
\end{tcolorbox}

%(Q23, Q24, Q25)
\subsection*{\RQC}
The previous RQ provided insight into where developers turn for assistance in securing their apps. This RQ seeks to understand the effectiveness of current educational materials in mobile app security. Additionally, it aims to gather recommendations from the developer community on improving mobile app security practices. We answer this RQ by examining the responses to survey questions \#21, \#22, and \#23. 

Looking at the responses to survey question \#21, the majority of the participants, 81 out of 137 (59.12\%) answered ``No'', indicating that the learning materials they used did not adequately cover the details needed to build secure mobile apps. The second most common response was ``Maybe'', with 39 participants (28.47\%), while 17 (12.41\%) answered ``Yes''.

Moving on, participants who answered ``No'' or ``Maybe'', were asked to share their thoughts about the lack of instruction on securing apps in a free-text format (survey question \#22). A review of these responses yields the following themes:
\begin{itemize}
    \item \textbf{Focus on Basic Development.} Security is often not a focus or priority in the learning materials, which instead emphasize basic app development concepts, UI/UX, and functionality (e.g., ``\textit{Most of the online materials are more focused on the UI design}'').
    \item \textbf{Need for Specialized Courses.} Several participants noted that app security is viewed as an advanced or separate topic, not typically included in introductory or beginner-level learning resources (e.g., ``\textit{If you wanna learn about security you need to search for a specific course about it}'').
    \item \textbf{On The Job Learning.} Security knowledge is often gained through practical experience, once developers start working on real projects or from more experienced colleagues (e.g., ``\textit{When I started working on real projects my mentor at the company started to suggest me best practices}'').
    \item \textbf{Security as a Secondary Concern.} In certain instances more priority is often assigned to get developers up to speed with basic app development as quickly as possible, sidelining security topics (e.g., ``\textit{Developing mobile apps was just starting so security was not a priority}''.)
    \item \textbf{Reliance on Platform-Specific Security Features.} Some participants, especially those using iOS, mentioned relying on the platform's security features and restrictions, which could reduce the need for separate security instructions (e.g., ``\textit{On iOS, by nature of the platform you are forced to be aware of certain security features}'').
    \item \textbf{Outdated or Incomplete Materials.} In some cases, the security content is outdated, incomplete, or lacks practical guidance on implementation (e.g., ``\textit{The documentation didn’t explain certain specific topics}'').
\end{itemize}

Next, participants were given an opportunity to answer a free-text question offering advice on mobile app security to fellow app developers (survey question \#23). A total of 128 participants responded to the question. An analysis of the responses yields the following themes:
\begin{itemize}
    \item \textbf{Continuous Learning.} Developers should proactively improve their security knowledge. This includes staying updated with official documentation, participating in community discussions, taking courses, and seeking out new information on security threats and solutions (e.g., ``\textit{Learn and read a lot about this because this is very important}'').
    \item \textbf{Best Practices and Standards.} Developers should adhere to established security guidelines and best practices in mobile app security. This includes secure coding practices, input validation, and staying updated with the latest security guidelines (e.g., ``\textit{Read OWASP, MASTG  and security guide for your platform}'').  
    \item \textbf{Proactive Security Integration.} It is crucial for developers to integrate security from the start of app development rather than treating it as an afterthought. Developers should find a balance between implementing strong security measures and maintaining a good user experience, and should proactively consider app security at every phase of the development process (e.g., ``\textit{Don't postpone it till the end of development, implement it continuously, to not waste lots of time}'').
    \item \textbf{Trusted Libraries and Security Tools.} Developers should utilize reliable, well-maintained, and up-to-date libraries. Tools like Proguard, R8, and Fortify are recommended for code obfuscation and vulnerability scanning (e.g., ``\textit{Try to use at least R8 and Proguard}'').
     \item \textbf{Data Protection.} It is essential that user data is protected through encryption, secure storage, and secure communication protocols (e.g., ``\textit{Encrypt the data and make sure your private keys are saved securely}''). 
     \item \textbf{Involve Security Professionals.} It is recommended that developers/organizations collaborate with or hire security professionals to ensure comprehensive app security. Developers can gain valuable insights and assistance in building secure applications by working with cybersecurity professionals (e.g., ``\textit{if possible, hire experts to audit everything from time to time}'').
     \item \textbf{Hands-on Experience.} Gaining hands-on experience through personal projects and thinking like a hacker helps developers better understand and implement security features (e.g., ``\textit{Try to apply your learned knowledge in a real or hobby project and test It}'').
      \item \textbf{Continuous Maintenance and Testing.} Developers should regularly update their apps and their dependencies to address security vulnerabilities and conduct thorough testing and validation of security measures, including vulnerability and penetration testing (e.g., ``\textit{test every part of your code}''). 
      \item \textbf{Knowledge Sharing.} Developers should engage with the mobile development community and share their knowledge and experience with mobile app security (e.g., ``\textit{Write and share new findings, there could be some exploits that others might not be aware}'').  
\end{itemize}

It should be noted that, as survey questions \#22 and \#23 are open-ended, some responses contain multiple themes.

\begin{tcolorbox}[top=0.5pt,bottom=0.5pt,left=1pt,right=1pt]
\textbf{RQ3 Summary.}
Many developers find that existing learning materials do not adequately prepare them to build secure apps. They point to several reasons, including the emphasis on basic app development, treating security as an advanced topic, and the necessity of gaining knowledge through practical experience, among others. Recommendations for developers regarding app security include following best practices, integrating security from the beginning, and using trusted tools and libraries.
\end{tcolorbox}

%(Q26)
The final survey question (\#24) was an optional free-text question that allowed participants to share anything else about their experience with mobile app security. There were 57 responses to this question. Although the participants had experience in mobile app development, some acknowledged they had limited knowledge or experience in securing mobile apps, and emphasized the need for more educational resources in this area  (e.g., ``\textit{I wish I could learn more, either a common resource or a well-known documentation hub for all of security}''). Participants also mentioned that security requirements depend on the app's purpose and context. Some of them also noted that security was not a concern for some of their apps (e.g., ``\textit{It depends on the type of app you are into. Some apps just don't need much to be secure}''). Additionally, participants highlighted the use of established libraries and security tools.

\section{Discussion}
\label{Section:discussion}
\rev{In this study, we provide an updated perspective on mobile app security practices from a developer's viewpoint. We investigate the challenges and opportunities for improvement, and examine how developers prioritize security implementation alongside other development tasks. We also explore the effectiveness of learning resources and gather practical recommendations from experienced developers. Below, we discuss key aspects of our analysis and provide takeaways to enhance mobile app security practices.}

\noindent\textbf{Disconnect Between Security Importance and Developer Preparedness.} Based on our findings from RQ1.1, it is clear that developers regard security as highly important. However, RQ3 indicates that current training materials do not sufficiently provide developers with the necessary security knowledge. This aligns with studies by Aljedaani et al. \cite{Aljedaani2020} and Balebako et al. \cite{Balebako2014}, which also demonstrates that app developers have inadequate security knowledge. This discrepancy highlights the disconnect between the acknowledged importance of security and the inadequate resources available to address it. Additionally, developers tend to learn about security through on-the-job experience, which can be problematic as it might lead to insecure and inconsistent security implementations, particularly from more junior developers, leaving the app vulnerable to potential vulnerabilities. \rev{To address these challenges, organizations should implement proactive onboarding processes involving comprehensive security training, including familiarizing with organizational security policies for new developers joining a project team. Further, continuous learning programs for employees will ensure that they are consistently updated on the latest security practices and techniques.}

\noindent\textbf{The Need for Improved App Dependency Practices.} While it is encouraging to see from RQ1.2 that most developers use trusted third-party libraries and ensure they are kept updated, it is concerning that only a minority of them rely on vulnerability scanning tools. From the developer recommendations in RQ3, the use of vulnerability scanning tools, such as Fortify, is highly encouraged. This contrast between common practices and expert recommendations suggests a need for greater awareness and adoption of automated security scanning in mobile app development. However, developers should not rely solely on these tools, as research indicates that while there are many security analysis tools available, their effectiveness in detecting known vulnerabilities is limited \cite{Ranganath2020}. This emphasizes the importance of a comprehensive approach to security that combines automated tools with other best practices and robust manual review processes.

\rev{
\noindent\textbf{Adopting Security-Driven Development.} Our RQ1.2 findings reveal that while many developers adhere to secure coding practices, only a minority actively engage in security testing. These findings are similar to those of Santos et al. \cite{Santos2020}. This highlights the necessity of implementing a holistic security-driven development approach, which involves integrating security considerations at every stage of the software development lifecycle, from initial design through implementation and testing \cite{olmsted2024security}. This would feature an emphasis on: incorporating security requirements and threat modeling into the early stages of development; adopting principles from Test-Driven Development by writing security tests before implementing features; establishing security acceptance criteria; including security tests in CI/CD pipelines, and performing tradeoff analysis \cite{kazman98} as part of design. By adopting a security-driven development approach, teams can ensure that security is not overlooked but is rather an integral part of the development process, which helps in identifying vulnerabilities early in the development lifecycle.}

\noindent\textbf{Organizational Security Resource Center.} Aligning with Aljedaani et al. \cite{Aljedaani2020}, our findings in RQ1.3 highlight that limited access to security resources, including personnel, is a major non-technical challenge for developers. Furthermore, based on our findings from RQ 2.2, it is evident that developers tend to depend on assistance from internal organizational resources.  The recommendations from RQ3 include involving security professionals, knowledge sharing, and continuous learning. Hence, we propose that organizations establish a comprehensive online Security Resource Center. This center would serve as a centralized hub, containing both industry-standard and organization-specific security resources, including best practices, security checklists, tools, company policies, training materials, sample secure code snippets, templates for security implementation, and contact information for in-house or external security experts. This easily accessible, curated, and up-to-date resource center would not only facilitate ongoing developer learning but can also ensure consistency in security practices across different projects and teams within the organization.

\section{Threats To Validity}
\label{Section:threats}
We selected participants for this study from LinkedIn because the platform enables us to review each individual's profile and confirm their experience in mobile app development. While there are other platforms such as Reddit and Discord, the anonymous nature of user profiles on these platforms makes it difficult to verify their expertise. Further, as this is not a funded study, we could not utilize platforms like Amazon Mechanical Turk and Prolific. Moreover, when evaluating user profiles we contacted only those who indicated involvement in mobile app development projects rather than just listing it as a skill. While this approach is useful for selecting participants for our study, it does have its limitations. Since user-created profiles cannot be verified, we may miss out on suitable candidates with incomplete or missing information. However, this threat of relying on self-reported information is a common issue in all survey-based studies.

To ensure the accuracy and reliability of the study findings, we gathered participant demographic information. While this data indicates that our participants are experienced professional app developers, there is a risk involved with relying on self-reported data. Selection bias is also a concern for this study,  as the developers who chose to respond may already consider security to be important. To mitigate this, we selected participants based on their app development experience without assessing their involvement in app/software security. However, it is possible that some level of selection bias may remain and could affect the generalizability of the findings. Moving on, the questions in our survey may not fully capture all aspects of mobile app security practices and challenges. Additionally, some survey questions consist of single or multi-choice responses, which might seem restrictive. To mitigate this risk, our survey also captures free-text responses to certain questions. Specifically, some single or multi-choice questions include an open-ended ``Other'' answer option, and the final survey question is an optional open-ended question, allowing participants to share any additional insights about app security. \rev{Furthermore, since our study did not include follow-up interviews with survey participants, we lack detailed rationales for some of their survey responses, which could have provided deeper insights into their decision-making processes and experiences. While not as comprehensive as in-depth interviews, the free-text response survey questions still offer valuable qualitative insight.}

To address researcher bias in interpreting qualitative (i.e., free-text) responses, two authors independently analyzed and categorized them, and then reached a consensus. 
Finally, we allowed anonymous survey responses to reduce bias, but the drawback is that we cannot verify the accuracy or legitimacy of the information provided, which could affect the overall reliability of the survey results.

\section{Conclusion \& Future Work}
\label{Section:conclusion}
Mobile apps play an essential part in people's lives, handling simple tasks like checking the weather and complex tasks like online banking. With most apps interacting with sensitive user information to achieve their desired functionality, developers must take precautions to secure their apps from issues that make them vulnerable to malicious attacks and data leaks.

In a survey of 137 professional app developers, we explored the common security features they use and the technical and non-technical challenges they face in securing their apps. Developers rely on various resources beyond Stack Overflow, but many feel that existing learning materials inadequately prepare them for real-world security challenges and offer recommendations for improving app security. 

\rev{Our future research involves conducting in-depth case studies with organizations that develop apps, to gain deeper insights into their security policies and best practices, furthering our understanding of this area. By examining diverse organizations across different industries and sizes, and by having opportunities to interview developers and explore topics in greater depth, we seek to identify diverse real-world examples of how companies effectively address mobile app security challenges and best practices. } 

\rev{These case studies will also analyze the type of mobile security education and training that developers on these projects receive, as well as the gaps in their knowledge and practices. This will help us understand the shortcomings of current training materials, explore the effectiveness of various training approaches, and propose improvements to curricula.}

\bibliographystyle{ieeetr}
\bibliography{main}
\end{document}

%% file: table_questionnaire.tex
\begin{table*}[t]
\centering
\caption{Below are the questions that are part of the survey. The questionnaire and the answer options for the single-choice and multi-choice, questions, are available at \cite{ProjectWebsite}.}
\vspace{-2mm}
\label{Table:SurveyQuestions}
\resizebox{\textwidth}{!}{%
\begin{tabular}{lp{0.7\linewidth}llp{0.2\linewidth}}
\toprule
\textbf{No.} &
  \textbf{Question} &
  \textbf{Type} &
  \textbf{Required} &
  \textbf{Notes} \\ \midrule\midrule
1 &
  Do you consent to participate in this study? &
  Yes/No &
  Yes &
  End survey for ``No'' response \\ \midrule
2 &
  Do you have experience in building and/or maintaining mobile apps? &
  Yes/No &
  Yes &
  End survey for ``No'' response \\ \midrule
3 &
  Which statement best describes your current employment status? &
  Single-Choice &
  Yes &
  Includes ``Other'' free-text option \\ \midrule
4 &
  How many years of general programming experience do you have? &
  Single-Choice &
  Yes &
   \\ \midrule
5 &
  How many years of mobile app development experience do you have? &
  Single-Choice &
  Yes &
   \\ \midrule
6 &
  To what extent is building mobile apps part of your job/employment duties? &
  Single-Choice &
  Yes &
   \\ \midrule
7 &
  How many mobile apps have you developed? &
  Single-Choice &
  Yes &
  Includes free-text option\\ \midrule
8 &
  Which mobile app distribution platforms do you utilize to share/distribute your apps? &
  Multi-Choice &
  Yes &
  Includes ``None'' \& ``Other'' free-text option\\ \midrule
9 &
  How did you initially learn mobile app development? &
  Single-Choice &
  Yes &
  Includes ``Other'' free-text option \\ \midrule
10 &
  How important is mobile app security to you when developing a mobile app? &
  Single-Choice &
  Yes &
   \\ \midrule
11 &
  What are the security features that you frequently implement in your mobile apps? &
  Multi-Choice &
  Yes &
  Includes ``Other'' free-text option \\ \midrule
12 &
  What steps do you take to ensure that your mobile app is not vulnerable to common mobile app security risks? &
  Multi-Choice &
  Yes &
  Includes ``Other'' free-text option \\ \midrule
13 &
  What steps do you take to ensure that third-party libraries and frameworks used in your mobile app are secure? &
  Multi-Choice &
  Yes &
  Includes ``Other'' free-text option \\ \midrule
14 &
  What are the frequent non-technical challenges you face in making your mobile apps secure? &
  Multi-Choice &
  Yes &
  Includes ``Other'' free-text option \\ \midrule
15 &
  What are the frequent technical challenges you face in making your mobile apps secure? &
  Multi-Choice &
  Yes &
  Includes ``Other'' free-text option \\ \midrule
16 &
  How frequently do you utilize Stack Overflow for help with mobile app security topics? & 
  Single-Choice &
  Yes &
   \\ \midrule
17 &
  How often do you find helpful information on Stack Overflow regarding mobile app security-related topics? & 
  Single-Choice &
  Yes &
   \\ \midrule
18 &
  How frequently do you use Stack Overflow to ask questions about mobile app security topics? & 
  Single-Choice &
  Yes &
   \\ \midrule
19 &
  How frequently do you answer a question on Stack Overflow related to mobile app security? &  
  Single-Choice &
  Yes &
   \\ \midrule
20 &
  What other information sources do you turn to for help with mobile security topics? & 
  Free Text &
  No &
   \\ \midrule
21 &
  Thinking back to when you were learning how to build mobile apps, did the learning material contain the necessary details on how to build secure mobile apps? & 
  Yes/No/Maybe &
  Yes &
   \\ \midrule
22 &
  Please share your thoughts about the lack of instruction on securing apps & 
  Free Text &
  Yes &
  Shown if Q21 is ``No'' or ``Maybe'' \\ \midrule
23 &
  What advice would you give to other mobile app developers who want to build secure apps? & 
  Free Text &
  No &
   \\ \midrule
24 &
  Is there anything else you want us to know regarding your experience with securing mobile apps? & 
  Free Text &
  No &
   \\ \bottomrule
\end{tabular}%
}
\end{table*}